\documentclass[10pt,a4paper,reqno]{amsart}
\usepackage{amsthm}
\usepackage{amsmath}
\usepackage{amsmath, amssymb}
\usepackage{type1cm}
\usepackage{cite}
\usepackage{bm}
\usepackage{cases}
\usepackage{here}
\usepackage[pdftex]{graphicx}
\theoremstyle{definition}

\newtheorem{prop}{Proposition}
\newtheorem{lemma}{Lemma}
\newtheorem{theorem}{Theorem}
\newtheorem{corollary}{Corollary}
\newcommand{\eqdef}{\stackrel{\mathrm{def}}{=}}
\DeclareMathOperator*{\esssup}{ess\,sup}
\begin{document} 
\title{Improved Chebyshev inequality: new probability bounds with known supremum of PDF}
\author{Tomohiro Nishiyama}
\begin{abstract}
In this paper, we derive new probability bounds for Chebyshev's inequality if the supremum of the probability density function is known.
This result holds for one-dimensional or multivariate continuous probability distributions with finite mean and variance (covariance matrix).
We also show that the similar result holds for specific discrete probability distributions.
\\

\end{abstract}
\date{}
\maketitle
\bibliographystyle{plain}
\section{Introduction}
The Chebyshev's inequality is a fundamental result in the field of probability theory and give probability bounds to a wide class of probability distributions.
If $X$ is a random variable with finite mean $\mu$ and finite non-zero variance $\sigma^2$, the Chebyshev inequality is as follows.\\

For any $\epsilon> 0$, 
\begin{equation}
\label{Chebyshev_inequaltiy}
\mathrm{Pr}\biggl(\frac{(X-\mu)^2}{\sigma^2}\geq \epsilon^2\biggr)\leq \frac{1}{\epsilon^2},
\end{equation}
where $\mathrm{Pr}$(A) is a probability on a set $A$.

There are several extensions to improve the sharpness of bounds (see e.g. \cite{savage1961probability}).
The well-known improved inequality is the one-sided (one-tailed) Chebyshev's inequality.
\begin{align}
\mathrm{Pr}\biggl(\frac{X-\mu}{\sigma}\geq \epsilon\biggr)\leq \frac{1}{1+\epsilon^2}
\end{align}

There are also several extensions to the multivariate case (see e.g. \cite{marshall1960multivariate} and references therein).
Recently, Chen has proved the n-dimensional multivariate Chebyshev's inequality\cite{chen2007new,navarro2016very}.
\begin{equation}
\label{multi_Chebyshev_inequaltiy}
\mathrm{Pr}((\bm{X}-\mu)^T\Sigma^{-1}(\bm{X}-\mu)\geq\epsilon^2)\leq\frac{n}{\epsilon^2},
\end{equation}
where $\bm{x}^T$ denotes the transpose of $\bm{x}$ and $\Sigma$ denotes the covariance matrix.

The classical and multivariate Chebyshev's inequality ((\ref{Chebyshev_inequaltiy}) and   (\ref{multi_Chebyshev_inequaltiy})) give the probability bounds on the set $D_{\epsilon}\eqdef \{\bm{X}\in\mathbb{R}^n | (\bm{X}-\mu)^T\Sigma^{-1}(\bm{X}-\mu)\geq \epsilon^2\}$.

In this paper, we improve the Chebyshev's inequality for continuous random variable and give new bounds on probability when the supremum of the probability density function is known in  $D_\epsilon$. The proof is based on the entropy upper bound with moment constraint\cite{conrad2004probability,marsiglietti2018lower}.
Then, we apply this inequality to specific discrete probability distribution.

\section{Main Results}
\textbf{Notation.}\\
\begin{itemize}
\item PDF: probability distribution function.\\
\item $E[\cdot]$: expected value.\\
\item$h(f)\eqdef -\int_{\mathbb{R}^n} f(\bm{x})\ln f(\bm{x})\mathrm{d}^nx$: differential entropy of PDF $f$.\\
\item$\|f\|_{\infty, A}\eqdef \esssup_{x\in A} f(x)$: uniform norm of PDF f on a set $A$.\\
  If $A=\mathbb{R}^n$, we omit $A$ on LHS.
\end{itemize}

\subsection{1-dimensional improved Chebyshev's inequality}
Before we derive new probability bounds, we show some lemmas.
\begin{prop}(entropy upper bound with fixed variance)\\
\label{variance_maximum_entropy}
Let $f$ be a PDF on $\mathbb{R}$ with variance $\sigma^2$.
Then, the following inequality holds.
\begin{align}
h(f)\leq \frac{1}{2}(1+\ln(2\pi\sigma^2))
\end{align}
Equality holds if and only if $f$ is normal distribution with variance $\sigma^2$.
\end{prop}
This Proposition is is shown as Theorem 3.2 in \cite{conrad2004probability}.

\begin{prop}(entropy upper bound with fixed $E[|x|])$\\
\label{mean_maximum_entropy}
Let $f$ be a PDF on $\mathbb{R}$ with fixed $\lambda=\int_{\mathbb{R}} |x|f(x)\mathrm{d}x$.
Then, the following inequality holds.
\begin{align}
h(f)\leq 1+\ln(2\lambda)
\end{align}
Equality holds if and only if $f$ is Laplace distribution with mean $0$ and variance $2\lambda^2$.
\end{prop}
This inequality is shown in \cite{marsiglietti2018lower}, and we can also prove in the same way as Theorem 5.2 in \cite{conrad2004probability}.
From Proposition \ref{variance_maximum_entropy} and \ref{mean_maximum_entropy}, we obtain the following lemma.
\begin{lemma}
\label{maxent_inequality}
Let $f$ be a PDF on $\mathbb{R}$ with finite mean and variance.
Then, the following inequalities hold.
\begin{align}
\label{x2_maxent}
\int_{\mathbb{R}} x^2f(x)\mathrm{d}x \geq \frac{1}{2\pi e}\exp(2h(f)) \\ 
\label{x_maxent}
\int_{\mathbb{R}} |x|f(x)\mathrm{d}x \geq \frac{1}{2e}\exp(h(f)) 
\end{align}
\end{lemma}
\noindent\textbf{Proof.}
Let $\mu$ be a mean and $\sigma^2$ be a variance.
By combining $\int_{\mathbb{R}} x^2f(x)\mathrm{d}x= \sigma^2 + \mu^2$ and Proposition \ref{variance_maximum_entropy}, we obtain (\ref{x2_maxent}).
Inequality (\ref{x_maxent}) is trivial.\\

\begin{lemma}(inequality for the uniform norm)\\
\label{Renyi_inequality}
Let $f$ be a PDF on $\mathbb{R}^n$.
Then, the following inequality holds.
\begin{align}
\exp(h(f))\geq \frac{1}{\|f\|_{\infty}}
\end{align}
\end{lemma}
\noindent\textbf{Proof.}
From $\|f\|_{\infty}\geq f(\bm{x})$ for $\bm{x}\in\mathbb{R}^n \mathrm{a.s.}$, we obtain $h(f)\geq -\int_{\mathbb{R}^n} f(\bm{x})\ln f(\bm{x})\mathrm{d}^nx\geq -\ln\|f\|_{\infty}$.
Exponentiating this inequality, we have the result. We can also obtain this result from the Renyi entropy inequality.\\

\begin{theorem}(1-dimensional improved Chebyshev's inequality)\\
\label{new_Chebyshev_inequality1}
Let $X\in\mathbb{R}$ be a continuous random variable with finite mean $\mu$ and variance $\sigma^2>0$, and $f$ be a PDF of $X$.
Let $D_{\epsilon}\eqdef \{X\in\mathbb{R} |\frac{(X-\mu)^2}{\sigma^2}\geq \epsilon^2\}$ and $m_{\epsilon}\eqdef \sigma\|f\|_{\infty,D_\epsilon}$.\\
For any $\epsilon>0$, the following inequality holds.
\begin{align}
\mathrm{Pr}(D_\epsilon)\leq \alpha m_\epsilon,
\end{align}
where $\alpha$ is a root of cubic equation $\frac{1}{2\pi e}x^3+\frac{\epsilon}{e}x^2+\epsilon^2 x-\frac{1}{m_\epsilon}=0$.
\end{theorem} 
\noindent\textbf{Proof.}
We change the variable as 
\begin{align}
Y\eqdef \frac{X-\mu}{\sigma}\\
\hat{f}(Y)\eqdef \sigma f(X).
\end{align}
For a random variable $Y$, $D_\epsilon$ can be written as $D_{\epsilon}=\{Y\in\mathbb{R} | Y^2\geq \epsilon^2\}$.
Then, we have 
\begin{align}
1&=\int_{\mathbb{R}} \frac{(x-\mu)^2}{\sigma^2} f(x)\mathrm{d}x=\int_{\mathbb{R}} y^2 \hat{f}(y)\mathrm{d}y\geq \int_{D_\epsilon} y^2\hat{f}(y)\mathrm{d}y.
\end{align}
Expanding $y^2=\epsilon^2+2\epsilon (|y|-\epsilon)+(|y|-\epsilon)^2$, we obtain
\begin{align}
\label{sigma_eval}
\int_{D_\epsilon} y^2\hat{f}(y)\mathrm{d}y=
\mathrm{Pr}(D_\epsilon)\epsilon^2+2\epsilon\int_{D_\epsilon} (|y|-\epsilon)\hat{f}(y)\mathrm{d}y+\int_{D_\epsilon} (|y|-\epsilon)^2\hat{f}(y) \mathrm{d}y
\end{align}
We define a new PDF as follows.
\begin{subnumcases}
{g(y)=}
\frac{1}{\mathrm{Pr}(D_\epsilon)}\hat{f}(y+\epsilon) & ($y>0$) \\
0 & ($y=0$) \\
\frac{1}{\mathrm{Pr}(D_\epsilon)}\hat{f}(y-\epsilon) & ($y<0)$
\end{subnumcases}
From definition, $g(y)$ satisfies $\int_{\mathbb{R}} g(y) \mathrm{d}y=1$.
By using $g(y)$, the equation (\ref{sigma_eval}) can be written as follows.
\begin{align}
1\geq \mathrm{Pr}(D_\epsilon)\biggl(\epsilon^2+2\epsilon\int_{\mathbb{R}} |y|g(y)\mathrm{d}y+\int_{\mathbb{R}} y^2g(y)\mathrm{d}y\biggr)
\end{align}
From Lemma \ref{maxent_inequality} and \ref{Renyi_inequality} , we have 
\begin{align}
1\geq \mathrm{Pr}(D_\epsilon)\biggl(\epsilon^2+\frac{\epsilon}{e}\exp(h(g))+\frac{1}{2\pi e}\exp(2h(g))\biggr)\geq \mathrm{Pr}(D_\epsilon)\biggl(\epsilon^2+\frac{\epsilon}{e\|g\|_\infty}+\frac{1}{2\pi e\|g\|_\infty^2}\biggr)
\end{align}
From the definition of $g(y)$ and $\hat{f}$, the equation $\|g\|_\infty=\frac{1}{\mathrm{Pr}(D_\epsilon)}\|\hat{f}\|_{\infty, D_\epsilon}=\frac{\sigma}{\mathrm{Pr}(D_\epsilon)}\|f\|_{\infty,D_\epsilon}$ holds.
Hence, we obtain
\begin{align}
\label{Pr_inequality}
1\geq \mathrm{Pr}(D_\epsilon)\epsilon^2+\frac{\epsilon\mathrm{Pr}(D_\epsilon)^2}{e\sigma\|f\|_{\infty,D_\epsilon}}+\frac{\mathrm{Pr}(D_\epsilon)^3}{2\pi e\sigma^2\|f\|_{\infty,D_\epsilon}^2}
\end{align}
By putting $m_\epsilon=\sigma\|f\|_{\infty,D_\epsilon}$, $t\eqdef \frac{\mathrm{Pr}(D_\epsilon)}{m_\epsilon}$ satisfies
\begin{align}
\frac{1}{2\pi e}t^3+\frac{\epsilon}{e} t^2 + \epsilon^2 t - \frac{1}{m_\epsilon}\leq 0
\end{align}
By putting $T(x)= \frac{1}{2\pi e}x^3+\frac{\epsilon}{e} x^2 + \epsilon^2 x - \frac{1}{m_\epsilon}$, we find $T(x)$ is a monotonically increasing function.
Hence, we obtain $t\leq \alpha$, where $\alpha$ satisfies $T(\alpha)=0$.
From the definition of $t$, we have the result.
\begin{figure}[H]
 \begin{center}
  \includegraphics[width=100mm, height = 70mm]{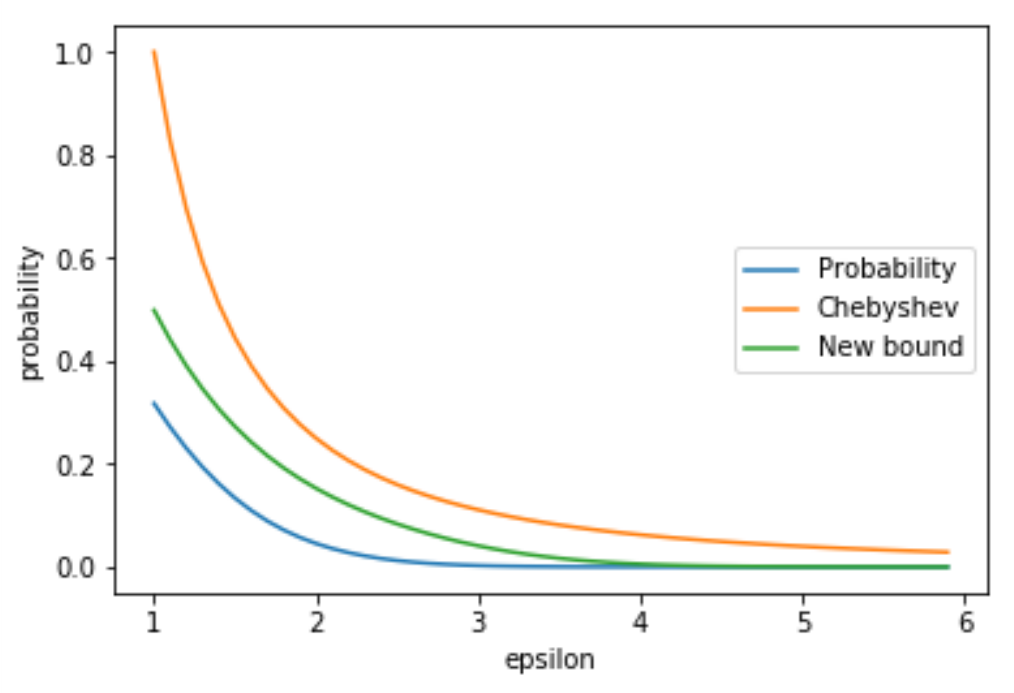}
 \end{center}
 \caption{Plots of actual probability and probability bounds for normal distribution. }
 \label{fig:one}
\end{figure}

In Figure 1, ''Probability'' denotes actual probability on the set $D_{\epsilon}\eqdef \{X\in\mathbb{R} |\frac{(X-\mu)^2}{\sigma^2}\geq \epsilon^2\}$,  ''Chebyshev'' denotes the result of Chebyshev's inequality and ''New bound'' denotes the result of Theorem \ref{new_Chebyshev_inequality1}.
\begin{corollary}
\label{new_Chebyshev_inequality2}
Let $X\in\mathbb{R}$ be a continuous random variable with finite mean $\mu$ and variance $\sigma^2$, and $f$ be a PDF of $X$.\\
Let $D_{\epsilon}\eqdef \{X\in\mathbb{R} |\frac{(X-\mu)^2}{\sigma^2}\geq \epsilon^2\}$ and  $m_{\epsilon}\eqdef \sigma\|f\|_{\infty,D_\epsilon}$.\\
For any $\epsilon>0$, the following inequality holds.
\begin{align}
\mathrm{Pr}(D_\epsilon)\leq \min(\frac{1}{\epsilon^2}, {\bigl(\frac{e}{\epsilon}\bigr)}^{\frac{1}{2}}{m_\epsilon}^{\frac{1}{2}}, {(2\pi e)}^{\frac{1}{3}}{m_\epsilon}^{\frac{2}{3}})
\end{align}
\end{corollary}
This inequality includes the Chebyshev's inequality.\\
\noindent\textbf{Proof.}
From (\ref{Pr_inequality}), we obtain
\begin{align}
\mathrm{Pr}(D_\epsilon)\epsilon^2\leq 1\\
\frac{\epsilon\mathrm{Pr}(D_\epsilon)^2}{em_\epsilon}\leq 1\\
\frac{\mathrm{Pr}(D_\epsilon)^3}{2\pi em_\epsilon^2}\leq 1.
\end{align}
From these inequalities, the result follows.
\subsection{Multivariate improved Chebyshev's inequality}
\begin{prop}(entropy upper bound with fixed covariance matrix)\\
\label{multivariate_variance_maximum_entropy}
Let $\bm{X}\in\mathbb{R}^n$ be a continuous random vector with  covariance matrix $\Sigma$, and $f$ be a PDF of $\bm{X}$.\\
Then, the following inequality holds.
\begin{align}
h(f)\leq \frac{1}{2}(n+\ln({(2\pi)}^n\det\Sigma))
\end{align}
Equality holds if and only if $f$ is n-dimensional normal distribution with covariance matrix $\Sigma$.
\end{prop}
This Proposition is shown as Theorem 5.5 in \cite{conrad2004probability}.
\begin{lemma}
\label{multivariate_maxent_inequality2}
Let $\bm{X}\in\mathbb{R}^n$ be a continuous random vector with  covariance matrix $\Sigma$, and $f$ be a PDF of $\bm{X}$.
Then, the following inequality holds.
\begin{align}
\frac{\mathrm{Tr}\Sigma}{n}\geq \frac{1}{2\pi e}\exp(\frac{2}{n}h(f))
\end{align}
\end{lemma}
By combining the positive definite matrix inequality (AM-GM inequality) $\frac{\mathrm{Tr}\Sigma}{n}\geq {(\det\Sigma)}^{\frac{1}{n}}$ and Proposition \ref{multivariate_variance_maximum_entropy}, we get the result.
\begin{theorem}(Multivariate improved Chebyshev's inequality)\\
Let $\bm{X}\in\mathbb{R}^n$ be a continuous random vector with  covariance matrix $\Sigma$, and $f$ be a PDF of $\bm{X}$.\\
Let $D_{\epsilon}\eqdef \{\bm{X}\in\mathbb{R}^n | (\bm{X}-\mu)^T\Sigma^{-1}(\bm{X}-\mu)\geq \epsilon^2\}$ and $m_{\epsilon}\eqdef{(\det\Sigma)}^{\frac{1}{2}}\|f\|_{\infty,D_\epsilon}$.\\
For any $\epsilon>0$, the following inequality holds.
\begin{align}
\mathrm{Pr}(D_\epsilon)\leq \min(\frac{n}{\epsilon^2}, {(2\pi e)}^{\frac{n}{n+2}}{m_\epsilon}^{\frac{2}{n+2}})
\end{align}
\end{theorem}
\noindent\textbf{Proof.}
As Chen have shown the inequality $\mathrm{Pr}(D_\epsilon)\leq \frac{n}{\epsilon^2}$, we prove the inequality $\mathrm{Pr}(D_\epsilon)\leq {(2\pi e)}^{\frac{1}{n+2}}{m_\epsilon}^{\frac{2}{n+2}}$.
We change the variable as 
\begin{align}
\bm{Y}\eqdef \Sigma^{-\frac{1}{2}}(\bm{X}-\mu)\\
\hat{f}(\bm{Y})\eqdef {(\det\Sigma)}^{\frac{1}{2}} f(\bm{X}).
\end{align}
For random variable $\bm{Y}$, $D_\epsilon$ can be written as $D_{\epsilon}=\{\bm{Y}\in\mathbb{R}^n | \sum_{i=1}^ny_i^2\geq \epsilon^2\}$.
We define a new PDF as 
\begin{subnumcases}
{g(\bm{y})=}
\frac{1}{\mathrm{Pr}(D_\epsilon)}\hat{f}(\bm{y}) & ($\bm{y}\in D_\epsilon$) \\
0 & ($\bm{y}\notin D_\epsilon$) 
\end{subnumcases}

\begin{align}
n&=\int_{\mathbb{R}^n} (\bm{X}-\mu)^T\Sigma^{-1}(\bm{X}-\mu)\mathrm{d}^nx=\int_{\mathbb{R}^n} \sum_{i=1}^ny_i^2 \hat{f}(\bm{y})\mathrm{d}^ny\geq \int_{D_\epsilon} \sum_{i=1}^ny_i^2 \hat{f}(\bm{y})\mathrm{d}^ny \\ \nonumber
&= \mathrm{Pr}(D_\epsilon)\int_{\mathbb{R}^n} \sum_{i=1}^ny_i^2 g(\bm{y})\mathrm{d}^ny\geq \mathrm{Pr}(D_\epsilon)\mathrm{Tr}\Sigma_g, 
\end{align}
where $\Sigma_g$ is the covariance matrix of $g(\bm{y})$ and we use $E[y_i^2]\geq \Sigma_{g,ii}$.
Combining this inequality, Lemma \ref{Renyi_inequality} and  Lemma \ref{multivariate_maxent_inequality2}, we have
\begin{align}
\label{multivariate_inequality}
1\geq \frac{\mathrm{Pr}(D_\epsilon)}{2\pi e}\exp(\frac{2}{n}h(g)) \geq  \frac{\mathrm{Pr}(D_\epsilon)}{2\pi e\|g\|_\infty^{\frac{2}{n}}}
\end{align}
From the definition of $g(\bm{y})$ and $\hat{f}(\bm{y})$, the equation $\|g\|_\infty=\frac{1}{\mathrm{Pr}(D_\epsilon)}\|\hat{f}\|_{\infty,D_\epsilon}=\frac{{(\det\Sigma)}^{\frac{1}{2}}}{\mathrm{Pr}(D_\epsilon)}\|f\|_{\infty,D_\epsilon}=\frac{m_\epsilon}{\mathrm{Pr}(D_\epsilon)}$ holds.
Substituting this equation to (\ref{multivariate_inequality}), we obtain the result.
\subsection{Application to specific discrete probability distributions}
We show examples of application of Theorem \ref{new_Chebyshev_inequality1} to specific discrete distributions.
\begin{theorem}
\label{discrete_inequality}
Let $Y\in\mathbb{Z}$ be a discrete random variable with finite mean $\mu$ and variance $\sigma^2$, and $p$ be a probability mass function.\\
Let $x_L\eqdef \mu+\frac{1}{2}-\epsilon\sigma_f$,  $x_R\eqdef \mu+\frac{1}{2}+\epsilon\sigma_f$ and  $\sigma_f^2\eqdef\sigma^2+\frac{1}{12}$.\\
Let $D_{\epsilon}\eqdef \{Y\in\mathbb{Z} |Y\leq\lfloor x_L\rfloor-1 \quad \mathrm{or} \quad Y\geq\lceil x_R\rceil\}$.\\
Let $M_{\epsilon}\eqdef \{Y\in\mathbb{Z} |Y\leq\lfloor x_L\rfloor\ \quad \mathrm{or} \quad Y\geq\lfloor x_R\rfloor\}$ and 
$m_{\epsilon}\eqdef \sigma_f\max_{k\in M_{\epsilon}} p(k)$.\\
For any $\epsilon > 0$, the following inequality holds.
\begin{align}
\mathrm{Pr}(D_\epsilon)\leq \alpha m_\epsilon,
\end{align}
where $\alpha$ is a root of cubic equation $\frac{1}{2\pi e}x^3+\frac{\epsilon}{e}x^2+\epsilon^2 x-\frac{1}{m_\epsilon}=0$.

\end{theorem}
\noindent\textbf{Proof.}\\
We define new PDF $f$ as follows.
\begin{align}
f(x)=p(\lfloor x\rfloor)
\end{align}
The mean and variance of $f$ are 
\begin{align}
E[X]=\int_{\mathbb{R}} xf(x)\mathrm{d}x=\sum_kp(k)\int_{k}^{k+1}x\mathrm{d}x=\mu+\frac{1}{2}\\
E[(X-E[X])^2]=\int_{\mathbb{R}} x^2f(x)\mathrm{d}x-\mu_f^2=\sum_kp(k)\int_{k}^{k+1}x^2\mathrm{d}x-\mu_f^2\\ \nonumber
=\sum_kk^2p(k)+\mu+\frac{1}{3}-\mu_f^2=\sigma^2+\frac{1}{12}=\sigma_f^2.
\end{align}
For the set $D^f_{\epsilon}\eqdef \{X\in\mathbb{R} |\frac{(X-\mu-\frac{1}{2})^2}{\sigma_f^2}\geq \epsilon^2\}$, we have
\begin{align}
\mathrm{Pr}(D_\epsilon)=\sum_{k\in D_\epsilon} p(k)=\sum_{k=-\infty}^{\lfloor x_L\rfloor -1}\int_{k}^{k+1}f(x)\mathrm{d}x + \int_{\lfloor x_L\rfloor}^{x_L}f(x)\mathrm{d}x\\ \nonumber
+ \sum_{k=\lceil x_R\rceil}^{ \infty}\int_{k}^{k+1}f(x)\mathrm{d}x + \int_{x_R}^{\lceil x_R\rceil}f(x)\mathrm{d}x\\ \nonumber
\leq \int_{D^f_\epsilon} f(x)\mathrm{d}x =\mathrm{Pr}_f(D^f_\epsilon),
\end{align}
where $\mathrm{Pr}_f$ denotes the probability of PDF $f$.
From Theorem \ref{new_Chebyshev_inequality1}, we have 
\begin{align}
\label{f_inequality}
\mathrm{Pr}(D_\epsilon)\leq \mathrm{Pr}_f(D^f_\epsilon) \leq \alpha m^f_\epsilon
\end{align}

where $m^f_\epsilon=\sigma_f\|f\|_{D^f_\epsilon}$ and $\alpha$ is a root of cubic equation $\frac{1}{2\pi e}x^3+\frac{\epsilon}{e}x^2+\epsilon^2 x-\frac{1}{m^f_\epsilon}=0$.
From the definition of $D^f_\epsilon$, we obtain $\|f\|_{D^f_\epsilon}=\max_{k\in M_{\epsilon}} p(k)$ and $m^f_\epsilon=m_\epsilon$.
Substituting $m^f_\epsilon=m_\epsilon$ to (\ref{f_inequality}), the result follows.

\begin{corollary}
\label{discrete_improved_Chebyshev}
Let $Y\in\mathbb{Z}$ be a discrete random variable with finite mean $\mu$ and variance $\sigma^2$, and $p$ be a probability mass function.\\
Let $x_L\eqdef \mu+\frac{1}{2}-\epsilon\sigma_f$,  $x_R\eqdef \mu+\frac{1}{2}+\epsilon\sigma_f$ and  $\sigma_f^2\eqdef\sigma^2+\frac{1}{12}$.\\
Let $D_{\epsilon}\eqdef \{Y\in\mathbb{Z} |\frac{(Y-\mu-\frac{1}{2})^2}{\sigma_f^2}\geq \epsilon^2\}$.\\
Let $M_{\epsilon}\eqdef \{Y\in\mathbb{Z} |Y\leq\lfloor x_L\rfloor\ \quad \mathrm{or} \quad Y\geq\lfloor x_R\rfloor\}$ and 
$m_{\epsilon}\eqdef \sigma_f\max_{k\in M_{\epsilon}} p(k)$.\\
For any $\epsilon > 0$, the following inequality holds.
\begin{align}
\mathrm{Pr}(D_\epsilon)\leq \alpha m_\epsilon+p(\lfloor x_L \rfloor),
\end{align}
where $\alpha$ is a root of cubic equation $\frac{1}{2\pi e}x^3+\frac{\epsilon}{e}x^2+\epsilon^2 x-\frac{1}{m_\epsilon}=0$.
\end{corollary}
\noindent\textbf{Proof.}
Let $D'_{\epsilon}\eqdef \{Y\in\mathbb{Z} | Y\leq\lfloor x_L\rfloor-1 \quad \mathrm{or} \quad Y\geq\lceil x_R\rceil\}$, we have
\begin{align}
\mathrm{Pr}(D_\epsilon)=\mathrm{Pr}(D'_\epsilon)+p(\lfloor x_L \rfloor)
\end{align}
From Theorem \ref{discrete_inequality}, the result follows.
\begin{figure}[H]
  \includegraphics[width=100mm, height = 70mm]{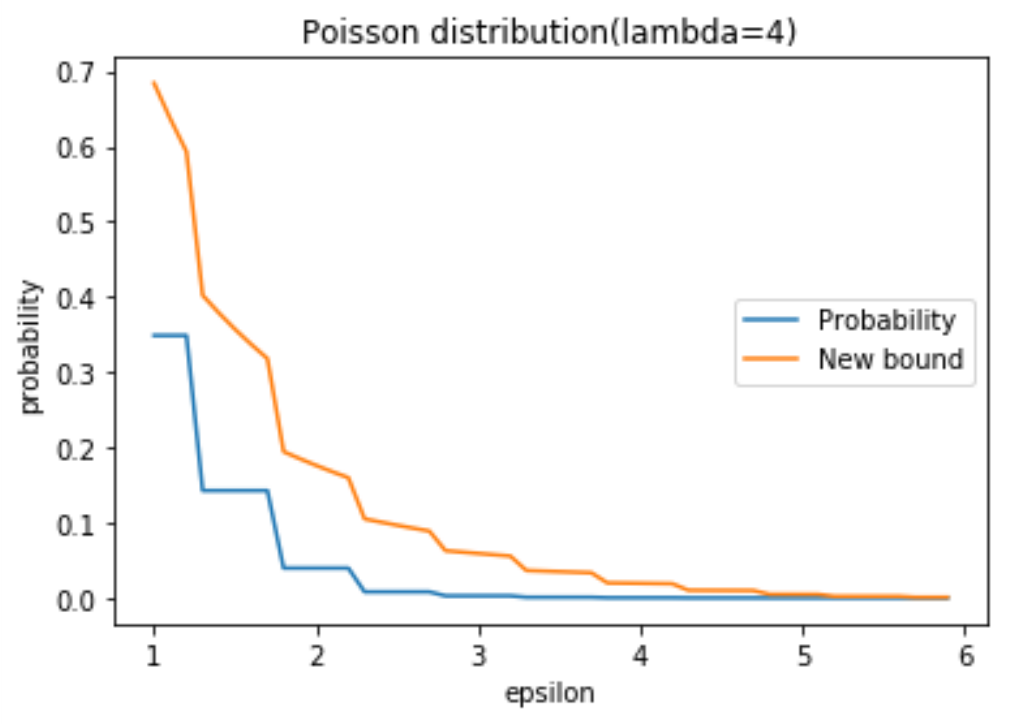}
 \caption{Plots of actual probability and probability bound for Poisson  distribution ($p(k,\lambda)=\frac{\lambda^k\exp(-\lambda)}{k!}$ and $\lambda=4$).}
 \label{fig:two}
\end{figure}

In Figure 2, ''Probability'' denotes actual probability on the set  $D_{\epsilon}\eqdef \{Y\in\mathbb{Z} |\frac{(Y-\mu-\frac{1}{2})^2}{\sigma_f^2}\geq \epsilon^2\}$ and ''New bound'' denotes the result of Corollary \ref{discrete_improved_Chebyshev}.
%
 
\section{Conclusion}
If the supremum of the probability density function is known, we have improved the Chebyshev's inequality for 1-dimensional or multivariate continuous probability distributions. We have also derived the similar inequality for specific discrete distributions by using improved Chebyshev's  inequality for continuous probability distributions.

Future works include expansion to more general discrete distributions and derivation of tighter bounds.
\bibliography{reference_v1}
\end{document}